\documentclass[aps,prl,twocolumn,superscriptaddress,email]{revtex4}  
\usepackage[pdftex]{graphicx}  
\usepackage{dcolumn}   
\usepackage{bm}        
\usepackage{amssymb}   

\usepackage{amsmath}
\usepackage[latin1]{inputenc}

\usepackage{color}

\newcommand{\ud}[1]{{#1^{\dagger}}}

\newcommand\Tr{\mathrm{Tr}}
\newcommand{\mean}[1]{\langle #1\rangle}

\begin{document}

\title{Effect of pure dephasing on the Jaynes-Cummings nonlinearities}
\author{A.~Gonzalez-Tudela}
\affiliation{F\'{\i}sica Te\'orica de la Materia Condensada, Universidad Aut\'onoma de Madrid, 28049, Spain.}
\author{E. del Valle}
\affiliation{School of Physics and Astronomy. University of Southampton, Southampton, SO171BJ, United Kingdom.}
\author{E. Cancellieri}
\affiliation{F\'{\i}sica Te\'orica de la Materia Condensada, Universidad Aut\'onoma de Madrid, 28049, Spain.}
\author{C. Tejedor}
\affiliation{F\'{\i}sica Te\'orica de la Materia Condensada, Universidad Aut\'onoma de Madrid, 28049, Spain.}
\author{D. Sanvitto}
\affiliation{F\'{\i}sica de Materiales, Universidad Aut\'onoma de Madrid, 28049, Spain.}
\author{F.P. Laussy}
\email[Corresponding author: ]{fabrice.laussy@gmail.com}
\affiliation{School of Physics and Astronomy. University of Southampton, Southampton, SO171BJ, United Kingdom.}

\date{\today}

\begin{abstract}
  We study the effect of pure dephasing on the strong-coupling between
  a quantum dot and the single mode of a microcavity in the nonlinear
  regime. We show that the photoluminescence spectrum of the system
  has a robust tendency to display triplet structures, instead of the
  expected Jaynes-Cummings pairs of doublets at the incommensurate
  frequencies~$\pm(\sqrt{n}\pm\sqrt{n-1})$ for integer~$n$. We show
  that current experimental works may already manifest signatures of
  single photon nonlinearities.
\end{abstract}

\maketitle 

Strong-coupling of quantum states---whereby bare modes vanish and
their quantum superpositions combining their properties take over---is
now commonplace in cavity Quantum Electrodynamics (QED)
physics. Following the seminal report with atoms~\cite{haroche89a}, it
is now realized with circuit QED~\cite{wallraff04a}, nano-mechanical
oscillators~\cite{arxiv_groeblacher09a} and, last but not least, with
quantum dots
(QD)~(see~\cite{hennessy07a,press07a,nomura08a,kistner08a,laucht09a,dousse09a}
for some recent reports, and references therein). Regarding the
latter, there have been recently rapid progresses in providing an
accurate quantitative description of the experiment, thanks to
theoretical efforts pointing towards specificities of the
semiconductor case, such as geometry of detection, self-consistent
states imposed by the incoherent pumping, etc.~\cite{cui06a,laussy08a}
These efforts culminated with the report by Laucht \emph{et
  al.}~\cite{laucht09b}, who provided a compelling global fit of
their experimental data. This approach allows, in the tradition of a
mature field of physics, to discard or validate a theoretical model by
confronting it statistically with the experiment. Laucht \emph{et al.}
have thus been able to bring out quantitatively which mechanisms
matter in the semiconductor strong-coupling case. Confirming the
suggestions of many previous
works~\cite{cui06a,naesby08a,yamaguchi08a,suffczynski09a}, they have
shown that a pure dephasing term is involved. Their analysis evidence
an exciton dephasing via interactions with phonons~\cite{borri05a} at
high temperatures, and with carriers outside the quantum
dot~\cite{favero07a} at high excitation power. In this Letter, we
shall not focus on the nature of the dephasing but take for granted
that it is not negligible.

The most appealing features of strong-coupling are at the quantum
level, when a few quanta of excitations rule the dynamics.  A
splitting at resonance is a tempting landmark of this regime, but is
not, being in no essential way different from the normal mode coupling
that is a classical feature of coupled oscillators~\cite{zhu90a}. To
evidence the quantum character of the coupling, photon-counting
experiments have been performed, reporting that only one quantum of
excitation couples the modes~\cite{hennessy07a,press07a}.  The next
step is to probe nonlinearities and witness their sensitivity at the
quantum level. 

The most basic and fundamental representation of the QD is that of a
two-level atomic-like system~\cite{michler00b}. Dressing this
fermionic system (i.e., coupling it strongly) with more than one
photon yields a splitting of~$2\sqrt{n}g$ when $n$ quanta are involved
($g$ is the interaction strength, we take $\hbar=1$).  Transitions
between these dressed states provide spectral lines at incommensurate
energies $\pm(\sqrt{n}\pm\sqrt{n-1})g$, which are a direct
manifestation of full-field quantization, as predicted by one of the
most important theoretical model of quantum physics, the
Jaynes-Cummings Hamiltonian~\cite{jaynes63a}. Evidencing these
nonlinearities is a chief goal of quantum optics. It has been
fulfilled with atoms~\cite{brune96a} and more recently with
superconducting circuits~\cite{fink08a}, but a direct spectral
signature remains elusive for semiconductor QDs, although compelling
indirect evidences have been reported~\cite{faraon08a,
  hennessy07a}. We have predicted that they could be observed with a
careful control (or lucky encounter) of the effective quantum
state~\cite{delvalle09a}. In this text, in the light of the importance
of pure dephasing in semiconductors, we revisit our claims taking it
into account. We show that due to dephasing, single-photon
nonlinearities manifest through a triplet at resonance in the
photoluminescence spectrum, rather than a quadruplet as
expected~previously.

The Jaynes-Cummings Hamiltonian, that describes the strong coupling of
a two-level QD with the single mode of a microcavity,
reads~$H=\omega_a a^\dag a +\omega_\sigma \sigma^\dag \sigma +g(
a^\dag \sigma +a \sigma^{\dag})$
%
with $a$ the photon annihilation operator (following Bose statistics)
and $\sigma$ the material excitation annihilation operator (following
Fermi statistics).  The two modes are coupled with interaction
strength $g$ and close enough to resonance (with small detuning
$\Delta=\omega_a -\omega_\sigma$) to allow for the rotating wave
approximation. A Liouvillian~$\mathcal{L}$ is used to describe the
system in the framework of a quantum dissipative master equation,
$\partial_t \rho=\cal{L}\rho$, taking into account decay~$\gamma_c$
and incoherent pumping~$P_c$, with~$c=a,\sigma$ (referred to as
\emph{cavity} and \emph{electronic} pumping,
respectively)~\cite{delvalle09a}.  Pure dephasing enters as an
additional source of decoherence~${\mathcal
  L}_{\gamma_\sigma^\phi}\rho$:~\cite{laucht09b}
\begin{multline}
  \label{eq:MonMay18154930BST2009}
  {\mathcal L} \rho= i [\rho,H]+\sum_{c =a,\sigma}\frac{\gamma_c}{2}(2 c \rho c^{\dag}-c^{\dag}c\rho-\rho c^{\dag}c)\\
  +\sum_{c=a,\sigma}\frac{P_c}{2}(2 \ud{c}\rho c-c\ud{c}\rho-\rho c\ud{c}) + {\mathcal L}_{\gamma_\sigma^\phi}\rho\,.
\end{multline}
The additional term~${\mathcal L}_{\gamma_\sigma^\phi}\rho$ originates
from high excitation powers or high temperatures. It disrupts
coherence without affecting directly the populations. It reads ${\cal
  L}_{\gamma_\sigma^\phi}\rho={\gamma_\sigma^\phi}(S_z\rho
S_z-\rho)$, where $S_{z}=\frac{1}{2}[\sigma^\dag,
\sigma]$~\cite{agarwal90a}.
%
%
As in our previous work~\cite{laussy08a, delvalle09a}, we recourse to
the quantum regression formula to compute the two-time average
$\mean{a^\dag(t)a(t+\tau)}$ which Fourier transform gives the
luminescence spectrum~$S_a$.  Following the same procedure as we
described before, we identify the tensor~$M$ that satisfies
$\Tr(C_{\{\eta\}}{\cal L}\Omega)=\sum_{\{\lambda\}}
M_{\{\substack{\eta\\\lambda}\}}\Tr(C_{\{\lambda\}}\Omega)$ for any
operator $\Omega$ in the basis of
operators~$C_{mn\mu\nu}=\ud{a}^ma^n\ud{\sigma}^\mu\sigma^\nu$.
The addition of pure dephasing only affects diagonal elements of~$M$
and for these, only when pertaining to phase coherence (when
$\mu\neq\nu$), where it is acting as a broadening:
\begin{multline*}
  \label{eq:MonMay18155354BST2009}
  M_{\substack{mn\mu\nu\\mn\mu\nu}}=i\omega_a(m-n)+i\omega_\sigma(\mu-\nu)\\-\frac{\gamma_a-P_a}2(m+n)-\frac{\gamma_\sigma+P_\sigma}2(\mu+\nu)-\frac{\gamma_\sigma^\phi}{2}(\mu-\nu)^2\,.
\end{multline*}
Other elements~$M_{\substack{mn\mu\nu\\pq\theta\vartheta}}$ are as given in
Ref.~\cite{delvalle09a}. In the following, we study the effect of
nonzero~$\gamma_\sigma^\phi$ on the spectral shape of the cavity
photoluminescence spectrum, under various cases of particular
experimental relevance.

\begin{figure}
  \centering
  \includegraphics[width=\linewidth]{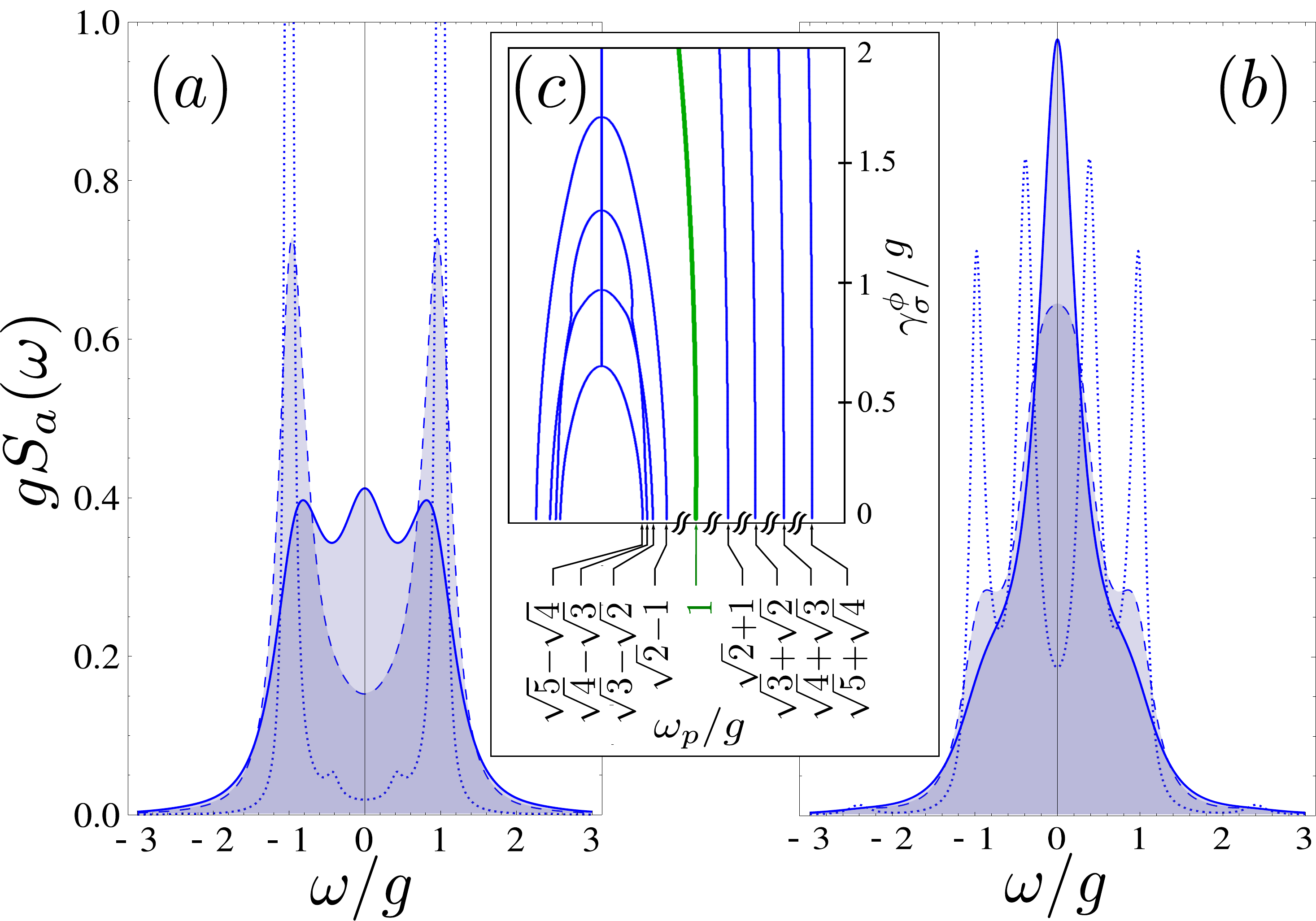}  
  \caption{(Colour online) Loss of the Jaynes-Cummings quadruplet and
    emergence of a triplet with dephasing, for a system well into
    strong coupling ($\gamma_a/g=0.1$ and
    $\gamma_\sigma/g=0.001$). Values of dephasing
    are~$\gamma_\sigma^\phi/g=0$ (dotted), $0.75$ (dashed) and $1.5$
    (solid). Panel~(a) [(b)] is for $P_\sigma/g=0.02$ [$0.1$]. In
    inset~(c), the dressed states resonances~$\omega_p/g$ for the
    parameters of~(b), showing the impact of dephasing on strong
    coupling: inner transitions melt into a common line. In thick
    green, the vacuum Rabi doublet maintains the satellites of the
    triplet.}
  \label{fig:SatJul4112134BST2009}
\end{figure}

Figure~\ref{fig:SatJul4112134BST2009} shows the impact of pure
dephasing on the most striking landmark of the Jaynes-Cummings
nonlinearities, namely, the multiplet structure that corresponds to
transitions between rungs of the Jaynes-Cummings
ladder~\cite{laussy09b}. We consider a system as realistic as possible
while still good enough to display them unambiguously.
A system with~$\gamma_a/g=0.1$ and~$\gamma_\sigma/g=0.001$, for
instance (which is still outside the reach of today's technology),
produces a Jaynes-Cummings fork (a quadruplet) at
resonance~\cite{delvalle09a}, as shown in dotted lines in
Fig.~\ref{fig:SatJul4112134BST2009} at lower~(a) and higher~(b)
pumpings.  With pure dephasing, the spectra evolve in both cases into
a triplet, with melting of the multiplet and emergence of a central
peak. The mechanism of this transition is revealed in the inset~(c),
where the dressed mode resonances~$\omega_p$ (in units of~$g$) are
shown as a function of the dephasing for the parameters of
panel~(b). These resonances correspond to transitions between the
eigenstates of the system, which are the dressed states (polaritons)
in the strong coupling regime, and the bare states (exciton and
photons) in the weak coupling regime. Dressed states up to five
photons are excited for the chosen parameters, and the characteristic
$\pm(\sqrt{n}\pm\sqrt{n-1})$ frequencies of the transitions between
rungs with such a square root splitting, are indicated at the bottom of
the figure (where~$\gamma_\sigma^\phi=0$). The system remains in
strong-coupling throughout, for all the states, as is evidenced from
the permanence of the outer resonances~$\pm(\sqrt{n}+\sqrt{n-1})$ for
all~$n\ge2$. The case~$n=1$ (in thick green) corresponds to the vacuum
Rabi splitting. Its position is only weakly perturbed by dephasing (as
are outer peaks). Inner transitions---when the decay links the same
type of states between two Jaynes-Cummings rungs (the two higher or
the two lower states)---are more significantly affected.  As shown in
the figure, these inner resonances, at~$\pm(\sqrt{n}-\sqrt{n-1})$
for~$n\ge2$, loose their splitting in succession with increasing
dephasing, the sooner the higher the excited state (i.e., the larger
the dressing). This loss of inner-splitting does not mean that the
system goes to weak coupling, but instead that the corresponding
transitions between dressed states are separated by a splitting
smaller than the uncertainty due to the dephasing, and as such, these
transitions overlap, thereby indeed providing the system with a new
common resonance, at the cavity mode. This transition is strong from
the accumulation of all the emissions of the system that were
previously split from each other (dephasing here acts like a quantum
eraser by providing an identical path to many previously
distinguishable paths). Further increasing dephasing eventually brings
the system into weak coupling, with collapse of the outer resonances
as well (not shown).  Triplets appear as a robust manifestation of
nonlinear strong coupling with dephasing: overlapping an emerging peak
at the cavity frequency with the vacuum Rabi doublet that produce
satellite peaks. In contrast to previously advanced
suggestions~\cite{hennessy07a,arxiv_ota09a}, our analysis shows that
this spectral structure is not attributable to loss of strong
coupling. It is also different from the triplet of Hughes and
Yao~\cite{hughes09a} that is due to interferences, and from the Mollow
triplet~\cite{mollow69a}, that arises in the classical limit of large
number of photons.  Instead, our triplet appears as a new regime at
the border of the quantum and classical regimes, with dephasing acting
as a smoothening agent (rather than a destructive
one~\cite{auffeves09a}).

\begin{figure}
  \centering
  \includegraphics[width=\linewidth]{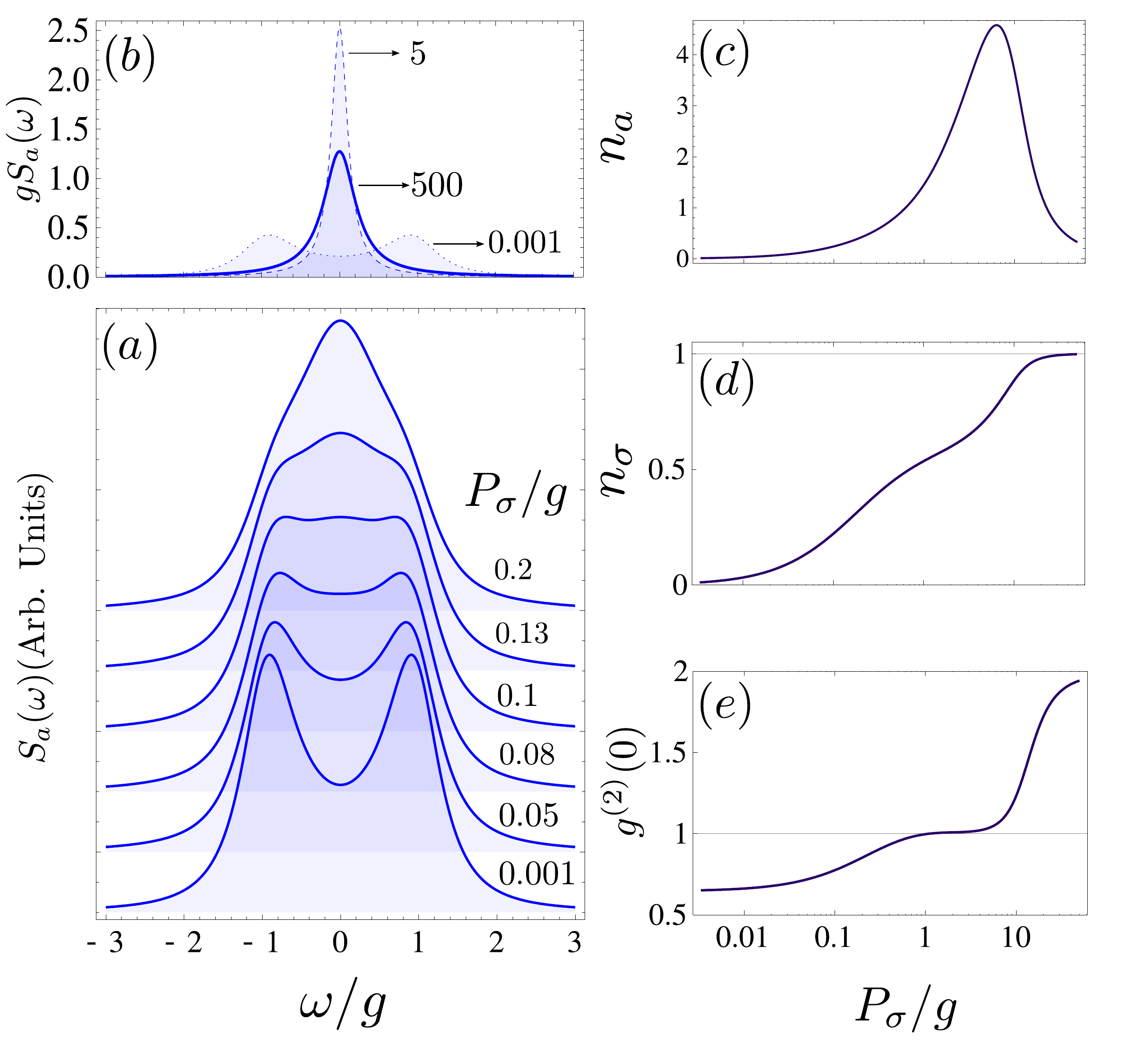}  
  \caption{Evolution of strong-coupling with increasing electronic
    pumping. Parameters are those of state of the art systems from the
    literature~\cite{nomura08a,arxiv_ota09a,laucht09a}: $g=120\mu$eV,
    $\gamma_a=38\mu$eV, $\gamma_\sigma=1\mu$eV,
    $\gamma_\sigma^\phi=g$, at resonance, $P_\sigma$ varying as
    indicated, without cavity pumping and with detector resolution
    of~46$\mu$eV. (a) The system evolves from the vacuum Rabi doublet
    into a triplet, much like the experiment of Ota \emph{et
      al.}~\cite{arxiv_ota09a} (that, however, is not strictly at
    resonance, making their triplet slightly better resolved at
    possibly smaller values of dephasing). (b) At much higher
    pumpings, the system goes to lasing then to quenching. (c-e) show
    these transitions in (c) the cavity population, (d) dot population
    and (e) $g^{(2)}(0)$.}
  \label{fig:SatJul4115249BST2009}
\end{figure}

Dephasing is not a parameter that is easy to control directly. In an
attempt to probe the nonlinearities of the system, a natural
experiment is to increase the pumping power, so as to populate more
the higher excited states. The evolution of the Rabi doublet with
increasing electronic pumping is shown on
Fig.~\ref{fig:SatJul4115249BST2009}(a), for parameters from
state-of-the art experimental systems (cf.~caption). In this case, a
triplet is also formed at resonance, but without any direct
manifestation of the Jaynes-Cumming quadruplets, owing to the poor
splitting to broadening ratio, even in the best systems available so
far.  The observation of this trend has been recently reported by Ota
\emph{et al.}~\cite{arxiv_ota09a}. We have indeed considered
parameters from this work to show that our effect is within the reach
of today technology. As the strong-coupling system is pushed more into
the nonlinear regime with pumping, a transition to lasing
occurs~\cite{delvalle09a, arxiv_nomura09a}. In panels~(b), we follow
this evolution in presence of dephasing, from the vacuum Rabi doublet
($P_\sigma/g\approx0.001$) towards a lasing single peak
($P_\sigma/g\approx5$) and eventually to a quenched system recovering
the bare cavity emission ($P_\sigma/g\approx500$). This is matched by
(c), the cavity population~$n_a$ (becoming~$>1$ with lasing and~$\ll1$
in the quantum/quenched regions), (d), the dot population (showing
population inversion with lasing and being empty or saturated in the
quantum/quenched regions) and, (e), the two-photon
coincidence~$g^{(2)}(0)$ (showing poissonian fluctuation at lasing,
and antibunching/bunching in the quantum/quenched regions).

\begin{figure}
  \centering
  \includegraphics[width=.65\linewidth]{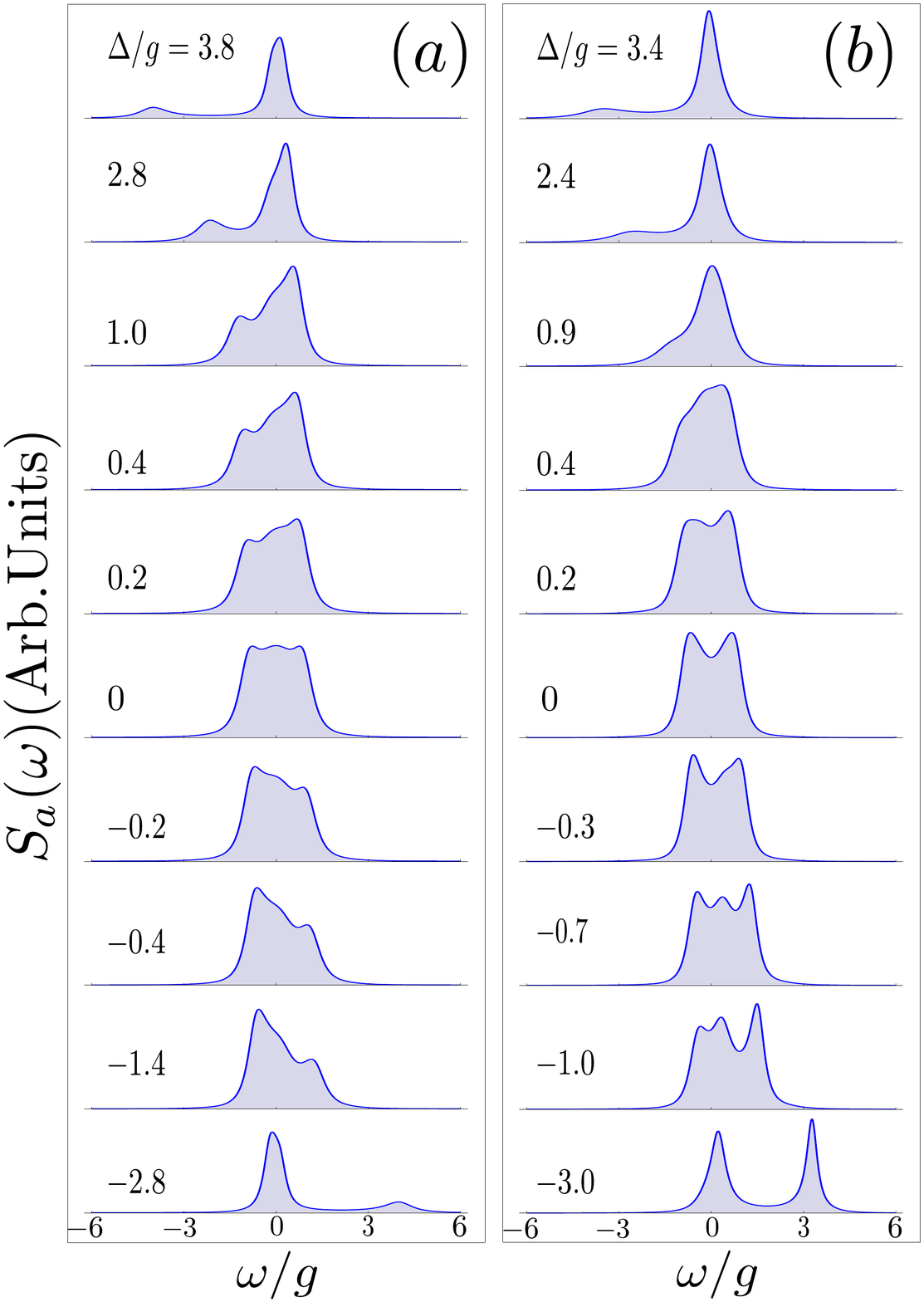}  
  \caption{Strong-coupling in the nonlinear regime in presence of
    dephasing as detuning is varied. Parameters are
    $\gamma_\sigma/g=0.001$ and $P_a/g=0.011$ for both panels, and for
    (a) [resp (b)], $\gamma_\sigma^\phi/g=1$ [sigmoid function
    of~$\Delta$], $\gamma_a/g=0.35$ [$0.5$] and $P_\sigma/g=0.1$
    [$0.3$].  Instead of the usual anticrossing, triplets are observed
    in slightly varying configurations: (a) A triplet is grown as the
    dot enters in resonance, much like the experiment of Hennessy
    \emph{et al.}~\cite{hennessy07a}. (b) As detuning varies with
    temperature, a triplet is observed \emph{out of resonance}, with
    an asymmetry with detuning caused by the temperature-dependent
    dephasing, much like the experiment of Sanvitto \emph{et
      al.}~\cite{arxiv_sanvitto06a}.}
  \label{fig:SatJul4115204BST2009}
\end{figure}

In Fig.~\ref{fig:SatJul4115204BST2009}, we display other
manifestations of nonlinearities in the luminescence spectrum of a
strongly-coupled QD/microcavity system, with the intent of showing the
wide range of phenomenologies that are accessible in different
configurations, as well as the strong similarities with other
experiments that have so far eluded a definite theoretical
explanation.  In these cases, we focus more on the similar general
behavior than on a tight numerical agreement with the experimental
values claimed in these works, although our parameters remain within
the possible margins for such systems (for instance, we have
considered an ideal detector in the cases of
Figs.~\ref{fig:SatJul4115204BST2009}, whereas we included the detector
resolution of Ref.~\cite{arxiv_ota09a} in our reproduction of their
experiment in Fig.~\ref{fig:SatJul4115249BST2009}). In the quest for
strong-coupling in semiconductors, one typically performs an
anticrossing experiment, where the dot and the cavity are brought to
resonance to exhibit level-repulsion (maintaining their line
splitting). Figure~\ref{fig:SatJul4115204BST2009} shows the situation
with detuning for two sets of parameters (cf.~caption).  In the first
case, (a), well identified dot and cavity emission lines approach in
the expected way but grow a central peak. This situation is similar to
the one reported by Hennessy \emph{et al.}~\cite{hennessy07a}. In the
second case, (b), a doublet is now produced at resonance and a triplet
is observed in its vicinity and only at negative detunings. This
situation is similar to the one reported by Sanvitto \emph{et
  al.}~\cite{arxiv_sanvitto06a} (that has remained unexplained---and
unpublished---so far). In the first case, dephasing is constant, the
cavity has a higher quality factor and electronic pumping is moderate.
The triplet then arises for the same reasons as those explained for
the phenomenology of Fig.~\ref{fig:SatJul4112134BST2009}. A slightly
better system (either from system parameters or with less dephasing)
would grow a quadruplet at resonance, if dephasing is indeed the cause
for the central peak in this case as well. These considerations match
the experimental situation of a single QD detuned from the cavity by a
thin-film condensation technique. In the second case, the experimental
situation varies in a few ways that would appear unimportant for the
physics investigated, but that turn out to produce very different
qualitative results: the dephasing has been correlated with the
detuning (with a sigmoid function, to reflect that detuning is tuned
with temperature), cavity photon has smaller lifetime and pumping is
much stronger. This results in the emergence of a triplet outside of
the resonance. In this later case, rather than superimposing a central
peak, the non-commensurable transitions placed
at~$\pm(\sqrt{n}-\sqrt{n-1})$ at resonance produce the multiplet out
of resonance, owing to their virtue of being stationary with
detuning~\cite{delvalle09a}. The dephasing here serves the purpose of
levelling the quadruplet predicted in Ref.~\cite{delvalle09a} for such
a structure at nonzero detuning, into a triplet.

In conclusion, we have shown that nonlinearities of the
Jaynes-Cummings Hamiltonian---the pinnacle of full-field quantization
in cavity Quantum Electrodynamics---have a robust tendency to manifest
as triplet structures in presence of a non-negligible dephasing (such
as is the case in semiconductors), rather than the expected
Jaynes-Cummings quadruplets with no emission at the cavity (central)
mode. We have shown that various parameters (corresponding to slightly
different experimental situations) result in strong qualitative
differences, such as observation of a triplet at---or out
of---resonance. Although Jaynes-Cummings nonlinearities in presence of
pure dephasing reproduce remarkably various experimental findings, on
the basis of a clear physical picture and with the expected
experimental parameters, a quantitative analysis is needed to bring a
definite proof that this effect is responsible for the observed
phenomenology. Experiments typically come with additional
complications of their own. For instance, a non-negligible drift in
detuning in Ota \emph{et al.}'s experiment is making their triplet
markedly more visible even at smaller values of the dephasing. Such a
compelling proof, however, is outside the scope of this Letter and a
challenge for the microcavity QED community at large.

La Caixa, Newton \& RyC programs, Spanish MEC (MAT2008-01555,
QOIT-CSD2006-00019) \& CAM (S-0505/ESP-0200), and EPSRC funds are
acknowledged.

\bibliography{Sci,jcdephasing}



\end{document}